\documentclass[useAMS,usenatbib]{mn2e}
\usepackage{epsfig}
\usepackage{amsfonts}
\usepackage[usenames]{color}
\usepackage[usenames,dvipsnames]{xcolor}
\usepackage{url}
\usepackage{subfigure}
\usepackage{times,graphicx,amsmath,amsfonts,amssymb,aas_macros,epstopdf,hyperref}
\usepackage[normalem]{ulem}
\usepackage[T1]{fontenc}
\usepackage{multirow}

\usepackage{graphicx}
\usepackage{amsmath}
\usepackage{astrobib_mnras2e}
\usepackage{times}
\usepackage{fixltx2e}
\usepackage[english]{babel}
\usepackage[T1]{fontenc}
\usepackage{ae,aecompl}
\usepackage{graphicx}	
\usepackage{amssymb}	
\usepackage{times,graphicx,amsmath,amsfonts,amssymb,aas_macros,epstopdf,hyperref}
\usepackage{epsfig}
\usepackage{amsmath,bm}
\usepackage{color}

\newcommand{\be}{\begin{equation}}
\newcommand{\ee}{\end{equation}}
\newcommand{\bs}{\begin{split} }
\newcommand{\es}{\end{split}}
\newcommand{\rewcom }{ }
\newcounter{defcounter}
\title{Optimal Redshift Weighting For Redshift Space Distortions}
\author[Ruggeri et al]{\parbox{\textwidth}{Rossana Ruggeri$^{1}$, Will J. Percival$^{1}$, H\'ector Gil-Mar\'in$^{1,2,3}$, Fangzhou Zhu$^{5}$, Gong-Bo Zhao$^{4,1}$, Yuting Wang$^{4,1}$}
\vspace*{15pt} \\
$^{1}$ Institute of Cosmology \& Gravitation, University of Portsmouth, Dennis Sciama Building, Portsmouth, PO1 3FX, UK\\
$^2$ Sorbonne Universit\'e, Institut Lagrange de Paris (ILP), 98 bis Boulevard Arago, 75014 Paris, France \\
$^3$ Laboratoire de Physique Nucl\'eaire et de Hautes Energies, Universit\'e Pierre et Marie Curie,
 Paris, France \\
$^{4}$ National Astronomy Observatories, Chinese Academy of Science, Beijing, 100012, P.R.China\\
$^{5}$ Dept. of Physics, Yale University, New Haven, CT 06511
}
\begin{document}

\maketitle
\begin{abstract}
  The low statistical errors on cosmological parameters promised by
  future galaxy surveys will only be realised with the development of
  new, fast, analysis methods that  reduce potential
  systematic problems to low levels. We present an efficient method
  for measuring the evolution of the growth of structure using Redshift Space
  Distortions (RSD), that removes the need to make measurements in
  redshift shells. We provide sets of galaxy-weights that cover a
  wide range in redshift, but are optimised to provide differential
  information about cosmological evolution. These are derived to
  optimally measure the coefficients of a parameterisation of the
  redshift-dependent matter density, which provides a framework to
  measure deviations from the concordance $\Lambda$CDM cosmology,
  allowing for deviations in both geometric and/or growth. We test the
  robustness of the weights by comparing with alternative
  schemes and investigate the impact of galaxy bias. We
  extend the results to measure the combined anisotropic Baryon
  Acoustic Oscillation (BAO) and RSD signals.
\end{abstract}
\begin{keywords}{cosmology: observations-
 large-scale structure of Universe - surveys - galaxies: statistics - cosmological parameters 
  }
\end{keywords}

\section{Introduction}
\label{sec:intro}

Forthcoming galaxy redshift surveys are motivated, to a large extent,
by obtaining galaxy clustering measurements to accurately quantify the
observed acceleration in the expansion of the Universe. It is to be
hoped that these observations will reveal insight into the physical
mechanism responsible for the cosmic acceleration, be it a new scalar
field currently contributing to the energy budget of the Universe as
Dark Energy, modification of gravitational laws on cosmological
scales, or an unknown alternative to the Standard Cosmological
Model.

Because the large-scale galaxy distribution is expected to follow a
Gaussian random field, for which the statistical information is fully
encoded in 2-point statistics, the central quantities in the analysis of
galaxy surveys are the Correlation Function and its Fourier-space
analogue, the Power Spectrum.  The observed projections of these
quantities encode significant cosmological information, including the
positions of the Baryonic Acoustic Oscillations (BAO), which can be
used as standard rulers to reconstruct the expansion history of the
Universe, \citep{bao15}.
 The statistics also
encode Redshift Space Distortions (RSD), which provide information
about the large-scale growth of cosmological structure, \citep{hamilton}.

The improvement in the statistical precision afforded by forthcoming
surveys including DESI, \citep{DESREF} and Euclid, \citep{laueucl2011} are
impressive, and will push at least an order of magnitude beyond
current measurements.  The improvement warrants a concerted effort to
improve the methods used to analyse these data, and recent key
developments include ``reconstruction'' to remove non-linear BAO
damping, \citep{recon}, and the development of fast methods to
measure the anisotropic clustering signal, \citep{bianchi2015,Scoccimarro2015}. 

One additional question to be answered is how best to combine future
data from different volumes within the surveys, without losing
\rewcom{information from galaxy pairs that span different bins, if using a binned approach}
 and to optimally recover the desired
signal. To deal with the first concern, we can make the transition
from splitting into redshift-bins, to instead adopting weights that
act to provide smoother windows on the data.

To optimise the weights,
we must consider two factors: the first concerns changes in the observational  efficiency  as a function of \rewcom{position on the sky and redshift}, and leads to weights that vary
as a function of observed galaxy density, \citep{FKP}, and bias , \citep{PVP}.
The second concerns the cosmological models that we wish to distinguish
between: \cite{FKP}, \cite{PVP} wished to optimally measure a power spectrum,
which was assumed to be fixed within a survey volume. If instead, we
wish to measure cosmological parameters that vary across a sample, for
example a quantity that evolves with redshift, then the weights must
additionally be optimised to measure this evolution.

\rewcom{In general, RSD measurements are made for a particular volume, presented as a single 
measurement at an effective redshift; if e.g. the growth factor varies in a non-linear way across the sample, the effective redshift is not a good approximation, by  contrast by weighting the sample we allow for variation in redshift of all the measured quantities. }

\cite{zhu2014} presented weights optimised for measuring the
distance-redshift relationship using the BAO signal. They considered a
second-order expansion of the distance-redshift relationship around a
fiducial cosmological model, and provided sets of weights for the
monopole and quadrupole moments of the correlation function (or power
spectrum) designed to optimally measure these parameters. In this
paper we extend this derivation to the measurement of Redshift-Space
Distortions, considering the weights required for these measurements,
and how they compare to the BAO-optimised weights.

The outline of the paper is as follows.  In Sec. \ref{sec:optweights}
we briefly go through the method of linear data compression and we
 underline the advantages related.
In Sec. \ref{sec:rsd} we present the cosmological model. 
In Sec. \ref{redweigh}
we  build to a derivation of the optimal weighting scheme for
redshift space distortion measurements parametrized with respect to
the matter energy density evolution in redshift, $\Omega_m(z)$. The
choice of $\Omega_m$ parametrization allows to easily extend the
results for more general clustering models which include both redshift
space distortion and Alcock \& Paczy\'nski effects (1979), AP.  In the
second part of the section we  compare them with other possible
weights optimized for RSD measurements and we  discuss our
assumption for linear bias  model.  In
Sec. \ref{biascompar} we  derive the generalisation of optimal
weights for RSD and AP test combined measurements. In
Sec. \ref{discussion} we  discuss our results and present potential
future improvements and  applications.

\section{Optimal Weights}  
 \label{sec:optweights}
The derivation of optimal weights is equivalent to the problem of
optimal data compression: we use the weights to reduce the number of
data points that need to be analysed to recover the cosmological
parameters. 
We will now review how to optimally linearly
compress our data, in the case of a covariance matrix known a priori,
as described in \citet{Tegmark97}. \rewcom{Further details on the Karhunen-Lo\`eve methods in e.g. \cite{1996vogeley} and \cite{pope2004}}.

Given the $n$-dimensional data-set ${\bf x}$, assumed to be Gaussian
distributed with mean ${\bf \mu}$ and covariance $C$, it can be linearly
compressed into a new data-set $y$,
\begin{equation}
y= \textbf{w}^T\textbf{x},
\end{equation}
where ${\bf w}$ is a $n$-dimensional vector of weights. The
measurement $y$ has mean ${\bf w}^T{\bf \mu}$ and variance
${\bf w}^TC{\bf w}$.

The Fisher information matrix $F$ is defined as the second derivative of
the logarithmic likelihood function $\mathcal{L} \equiv - \ln L$,
\begin{equation}  \label{eq:Fii}
  F_{ij} \equiv \left\langle \frac{\partial^2\mathcal{L}}{\partial \theta_i \partial \theta_j}\right\rangle,
\end{equation} 
for a set of parameters to be measured $\theta_i$. For a single
parameter $\theta_i$, 
\begin{equation} 
  F_{ii} =
  \frac{1}{2}\left(\frac{{\bf w}^T C_{,i}{\bf w}}{{\bf w}^T C {\bf w} } \right)^2+
  \frac{\left({\bf w}^T \mu_{,i} \right)^2}{ {\bf w}^T C {\bf w} },
\label{fishmat}
\end{equation}
where the index $,i $ denotes $\partial / \partial \theta_i $. Note that the normalisation of the weights is  arbitrary
The search for optimal weights is equivalent to maximising $F_{ii}$
with respect to ${\bf w}$. 

For a measurement of 2-point statistics from a galaxy survey, we
should consider that ${\bf x}$ is the \rewcom{arrays} formed by the 
measurements of the over-density squared, $\delta^2$ in configuration or Fourier space. Working in Fourier space, the covariance matrix $C$ of the power spectrum of the modes in the
absence of a survey window is diagonal; \rewcom{for each redshift slice with volume $dV$ and expected galaxy density $\overline{n}(r)$}, 
\begin{equation}
  C \sim \left(  P_{\rm fid} + 1/\overline{n}(r)  \right)^2 \dfrac{1}{dV},
\end{equation}
where we have made the assumption that around the likelihood maxima,
the power spectra $P_{fid}$ are drawn from a Gaussian distribution with fixed
covariance matrix, e.g. \cite{kalus15}. In this case, the first
term in Eq.~(\ref{fishmat}) vanishes. Maximising $F_{ii}$ with respect
to ${\bf w}$, we find the only non-trivial eigenvector to be
\begin{equation} 
  {\bf w}^T = C^{-1} {\bf \mu}_{,i},
\label{defw}
\end{equation}
and the new compressed data set reduces to 
\begin{equation}
  y =  {\bf \mu}_{,i}^T C^{-1}{\bf x}.
\end{equation}
Note that, to linear order, $y$ contains the same information as
${\bf x}$, which can be checked by substituting
${\bf w} = C^{-1} {\bf \mu}_{,i}$ in Eq.~(\ref{fishmat}), and seeing
that $F$ remains unchanged. Eq.~(\ref{defw}) forms the basis for our
derivation of optimal weights.
    
Eq.~(\ref{fishmat}) shows why it does not make sense to optimise
for the set of $\delta$ (as opposed to $\delta^2$). This is because,
although now the second term in Eq.~(\ref{fishmat}) now vanishes as $\langle \delta \rangle = 0$, the
resulting eigenvector equation derived from the first term shows that
there is no single set of optimal weights, even under the simplifying
assumption of a diagonal covariance matrix. We can still apply the
weights derived for $\delta^2$ to individual galaxies if we assume
that the scales upon which clustering is being measured are small with
respect to the cosmological changes that affect the relative
weights. We would then simply weight each galaxy (and the expected
density used to estimate $\delta$) by
${\bf w}_{\rm gal}=\sqrt{{\bf w}_{\delta^2}}$.

Note that the optimal set of weights given in Eq.~(\ref{defw}) depends
upon the derivatives of $\mu_{i} = P_{,i}  $. 
Consequently in the rest of the
paper we concentrate our analysis on the form of $P_{,i}$, which
directly gives the form for the weights. If $P_{,i}$ matches for
different measurements, then the optimal weights will also match.

The weights can be seen as a generalisation of the FKP weights
presented in \cite{FKP}.  The FKP weights are obtained by minimising
the fractional variance in the power under the assumption that
fluctuations are Gaussian and have the form
\begin{equation}
w_{\rm FKP}(r) = \frac{1}{1+ \overline{n}(r) P(k)};
\end{equation}
The weights defined by Eq.~(\ref{defw}) depend on the inverse of the
covariance matrix: assuming the redshift slices to be independent we
can invert the Covariance matrix for a chosen scale and recover the
FKP weights.

The cosmological model-dependent weights depend on the covariance
matrix assumed and on the derivative of the mean value of the model
with respect to the parameter that we want to estimate. Thus, once
these quantities are fixed, it is not trivial to adapt a particular
set of weights for different models. However, it would be very useful
to set up weights that can be applied to make different measurements:
both for computational reasons and in order to perform joint fits to
the data. In this work we will start by deriving weights to be applied
to RSD measurements and then we will broaden this to consider jointly
measuring the RSD and the AP effect. Comparing the set of weights in
these different situations shows whether it is likely that a single
set of weights can be used to make optimal measurements in both
situations.


\section{Cosmological Model}
\label{sec:rsd}

\subsection{Fiducial Cosmology}
\label{fidco}
The $\Lambda$CDM  scenario  predicts the nature of dark energy  as a cosmological constant with equation of state parameter $w = -1$ where the dynamical expansion of the Universe is specified by Friedmann equation
\begin{equation}
\frac{H_{\rm fid}^2 (z)}{ H_{0,fid}^2} = \Omega_{m,0, \rm fid}(1+ z)^3 + \Omega_{k,\rm fid} (1+ z)^2  + \Omega_{\Lambda,\rm fid}(z),
\label{fried}
\end{equation}
where the subscript  the ``$0$'' stands for quantities evaluated at $z= 0$ while ``$\rm fid$'' denotes   fiducial quantities.
With $\Omega_{\Lambda,\rm fid} $ dark energy density, $\Omega_{k,\rm fid}  = 1-
\Omega_{m,\rm fid} - \Omega_{\Lambda,\rm fid} $ curvature, and $H_0$ the
present-day Hubble parameter. We have 
\begin{equation}
\Omega_{m,\rm fid}(z) = \frac{\Omega_{m,0,\rm fid}(1+z)^3}{H_{\rm fid}^2(z)/H^2_{0,\rm fid}},
\label{fidomega}
\end{equation}
where $\Omega_{m,0}$ refers to energy density evaluated at $z = 0 $.
In a Friedmann-Roberston-Walker universe the solution for the linear
growth factor $D_{\rm fid}(z)$ and the dimensionless linear growth rate $f$
are given by
\begin{equation}\begin{split}  \label{eq:gf}
g_{\rm fid}(\Omega_{m,\rm fid}(z)
) & \equiv   \frac{D_{\rm fid}(z)}{a}  \\&=
\frac{5\Omega_{m,\rm fid}(z)H_{\rm fid}^3(z)}{2(1+z)^2} \int_z^\infty\mathrm{d}z'\frac{ (1+z')}{H_{\rm fid}^3(z')} 
\end{split}
\end{equation}
with scale factor  $a$; 
 \begin{equation} \begin{split}  \label{eq:ff}
 f_{\rm fid}(\Omega_{m,\rm fid}(z)
  )
  = &-1 -\frac{ \Omega_{m,\rm fid}(z)}{2} +\Omega_{\Lambda,\rm fid}(z) \\&+ \frac{5\Omega_{m,\rm fid}(z)}{2g_{\rm fid}(z)} .
\end{split}  \end{equation} 
Under the assumption of a flat universe i.e. $\Omega_{k,\rm fid} = 0 $,
we have  $\Omega_{\Lambda,\rm fid}(z) = 1- \Omega_{m,\rm fid}(z)$.

For the fiducial galaxy bias model we choose a simple \textit{ad hoc} functional form
as used in \cite{rassat2008}, which is approximately correct for the
$H\alpha $ galaxies to be observed by the Euclid survey,
\begin{equation}
b_{\rm fid} = \sqrt{1+ z}.
\label{eubias}
\end{equation}

\subsection{Parametrising deviations}
\label{sec:dz}

The derivation presented in Sec. 2 required us to define the parameters
that we wish to optimise measurement of. We wish to choose parameters
that allow us to measure deviations from the $\Lambda$CDM model. In
the absence of compelling alternative cosmological models, we choose
parameters that define an expansion in redshift of the cosmological
behaviour we wish to understand - in our case the structure growth
rate, and the expansion rate. Both of these can be modelled by
deviations in $\Omega_{m}(z)$ away from the fiducial model, and we
adopt this quantity as the redshift-evolving quantity that we wish to
understand, rather than the distance-redshift relation considered by
\citet{zhu2014}, which does not easily extend to structure growth
differences. We expand $\Omega_{m}(z)$ around the fiducial model as,
\begin{equation}
\dfrac{\Omega_m(z)}{ \Omega_{m,\rm fid}(z)}=  q_0 (1 + q_1y(z)  +\frac{1}{2} q_2 y(z)^2 ),
\label{espomega}
\end{equation}
we fix a pivot redshift $z_p$ within the survey redshift range and $y$
is defined as
$y(z)+ 1 \equiv \dfrac{\Omega_{m,\rm fid }(z) }{\Omega_{m,\rm fid}(z_p) } $; the
expansion parameters $q_0$, $q_1$, $q_2$ are obtained from
Eq. \ref{espomega} and its first and second derivatives evaluated at
$z_p$;
\begin{equation}\begin{split}
q_0 =& \dfrac{\Omega_m(z_p)}{\Omega_{m,\rm fid}(z_p) },\\
q_1 =& \dfrac{\Omega_{m,\rm fid}(z_p)}{\Omega_{m}(z_p) } \dfrac{ d \Omega_{m} /dz |_{z_p}}{d \Omega_{m,\rm fid}/ dz |_{z_p} } -1 ,\\
q_2 =& \bigg[ \frac{\Omega_{m,\rm fid}(z_p)}{\Omega_{m}(z_p)} \frac{d^2 \Omega_{m} }{dz^2}|_{z_p} - \frac{d^2\Omega_{m,\rm fid} }{dz^2}|_{z_p} \\
 &- \left. \left(    \dfrac{\Omega_{m,fid}(z_p)}{\Omega_{m}(z_p) } \dfrac{ d \Omega_{m} /dz |_{z_p}}{d \Omega_{m,\rm fid}/ dz |_{z_p} }  -1 \right )
 \left(\dfrac{2}{\Omega_{m,\rm fid}(z_p) }\right.  \right. \\
&  \left( \frac{d\Omega_{m,\rm fid}}{dz}\right)^2|_{z_p} 
  \left. + \frac{d^2\Omega_{m,\rm fid}}{dz^2}|_{z_p} \right) \bigg]\dfrac{\Omega_{m,\rm fid }(z_p)  }{ \left( d \Omega_{m,\rm fid}/ dz |_{z_p} \right)^2 }.
\end{split}
\end{equation}
The Hubble parameter with respect to $\Omega_m( z)$ is
\begin{equation}
  \frac{H^2( z)}{H^2_0} =\frac{\Omega_{m,0} (1+z)^3 }{\Omega_m( z)},
  \label{hubpar}
\end{equation}
where we have assumed that the dark matter equation of state is fixed,
\begin{equation}
  \mathcal{P}= w \rho; \; w =0 
\end{equation} 
with pressure $\mathcal{P}$ and matter density $\rho$. 

A broader range of models could be derived by perturbing the homogeneous solution of the
Einstein equations, but in this work we restrict ourselves to
deviations close to $\Lambda \rm CDM$ in the dark energy and curvature
components.  The $\Omega_m$ parametrisation allows for many deviations
from $\Lambda \rm CDM$: all the standard cosmological parameters can be
written in terms of the $q_i$ parameters e.g. if we want to allow for
modified gravity models we can parametrize the growth factor as a
function of $\Omega_m(z)$.  Alternatively, if we are studying the
deviations from a fiducial geometry we can parametrize the AP
parameters. We assume that Eqns.~(\ref{eq:gf}) \&~(\ref{eq:ff}) hold
for the perturbed $\Omega_m(z )$, which fix how the parameters
$q_i$ lead to deviations in the growth rate away from the fiducial
model.

\section{Redshift weighting assuming known distance-redshift relation} \label{redweigh}
In this section we derive a set of optimal weights  to  measure  $\Omega_m(z)$ and the bias when the distance-redshift relation is assumed known, i.e  we derive the weights for an observed power spectrum that contains only the distortion due to RSD and not due to AP effect.

We will discuss three different cases with different assumptions about  parameters assumed known: the first two aim to estimate  $\Omega_m(z)$  and the third the bias, parametrised by $b\sigma_8(z)$.
In \ref{allinfoweights} we consider the case in which the bias is known and fixed to a fiducial model; in this case we are considering that the information about $\Omega_m$ is coming from all the terms of the Power Spectrum multipoles. 
We discuss this set of weight with respect to  different fiducial models for the bias in order to test how knowing the bias would affect the results. 
However this first set of weights does not match the actual RSD measurements condition in which the bias is unknown.
We then consider in \ref{sec:weights_APfixed_fsig8} a case for RSD measurements where
we  assume the growth information is not coming from tangential power and the only information to be  considered comes from  $f \sigma_8$.
%
 Bias evolution plays an important role for clustering measurements
even if, in the redshift range of interest for future and current
surveys, $\Omega_m$ is significantly more sensitive to redshift than
bias, (see Sec. \ref{sec:rsd}).
 As our last case, in \ref{sec:weights_APfixed_bias},  we consider for completeness a set of weight to measure the bias relation as a function of redshift.

\subsection{Modelling the observed power spectrum}
\label{obpow}
For simplicity we adopt a linear model for the redshift-space
distortions, assume that we are working in the plane parallel
approximation, and assume a linear deterministic bias model so that
the power spectrum in redshift space, $P^s$ is related to the real
power spectrum $P$ by
\begin{equation}  \label{eq:pks}
P^s (\textbf{k} ) = (b+ f\mu^2_{\textbf{k} })^2 P(k)
\end{equation}
where $P(k)$ is the linear real space power spectrum and where $\mu_\textbf{k}
\equiv \hat{\textbf{z}} \cdot \hat{\textbf{k}}  $ is the cosine of the
angle between the wavevector $\textbf{k}$ and the line of sight
$ \hat{\textbf{z}}$, \citep{kaiser1987}. 

It is common to decompose $P^s$ into an orthonormal basis of Legendre
polynomials such that, in linear regime, the redshift power spectrum is well
described by its first three non-null moments: monopole $P_0$, quadrupole $P_2$
and hexadecapole $P_4$.
\begin{equation} 
  P^s(\textbf{k}) = \mathcal{P}_0(\mu_\textbf{k} )P_0(k) +
  \mathcal{P}_2(\mu_\textbf{k} )P_2(k) +
  \mathcal{P}_4(\mu_\textbf{k} )P_4(k), 
\end{equation}
related with $P(k)$ through  ,
\begin{equation} 
  P_{0}(k) = \left(b^2 +\frac{2}{3} b f +\frac{1}{5}
    f^2\right)P(k) \label{m}
  \end{equation},
\begin{equation} 
  P_{2}(k) = \left(\frac{4}{3}bf  + \frac{4}{7} f^2\right)
  P(k), \label{q} 
\end{equation}
\begin{equation} 
  P_{4}(k) = \left(\frac{8}{35}f^2 \right) P(k) \label{h}. 
\end{equation}

We normalise the power spectrum using the standard variance of the
galaxy distribution smoothed on scale $R = 8 h^{-1}$Mpc,
$\sigma_{8}(z)$, where 
\begin{equation}
  \sigma_{8}(z) =  \sigma_{8,0} D(z) =\sigma_{8,0} \frac{g(z)}{1+z} .
\end{equation}
This normalisation enters into Eq.~(\ref{eq:pks}) in a way that is
perfectly degenerate with $b$ and $f$, which could be replaced by new
parameters $(b\sigma_8)$ and $(f\sigma_8)$. 
\subsection{Optimal weights to measure $\Omega_m(z)$ assuming known bias} 
\label{allinfoweights}
We build optimal weights by taking the derivative of the power
spectrum model with respect to the parameters $q_i$, 
As discussed in section \ref{sec:optweights},  hereafter we only consider the component of these weights that varies with $P_{,i}$, which we denote for the monopole,
quadrupole and hexadecapole, respectively $w_0$, $w_2$,
$w_4$. For simplicity we refer to these as the ``weights'', but it is worth remembering that there is a missing inverse variance component. 
\begin{equation}
  w_{\ell, q_i} = \dfrac{\partial P_\ell}{\partial q_i}
  \label{ww}
\end{equation}
We explicitly write the redshift dependence of P on the $q_i$ parameters, 
so that the right side of Eq.~(\ref{ww}) becomes 
\begin{equation}
  \dfrac{\partial P_\ell }{\partial q_i} =   
    \dfrac{\partial P_\ell}{\partial f} \dfrac{\partial f}{\partial q_i}+   
    \dfrac{\partial P_\ell }{\partial \sigma_8} \dfrac{\partial \sigma_8 	}{ \partial q_i}  
\label{primweigh}
\end{equation}
This second term  assumes that we are recovering information from both radial and \rewcom{transverse modes}. 
This is true for the \rewcom{transverse} component if the bias is known perfectly.
We build  our set of weights as a function of redshift, for the
monopole we have
\begin{equation}
w_{0,i} = \left(\frac{2}{3} b + \frac{2}{5} f  \right)\sigma_8^2 \frac{\partial f}{\partial q_i}
  + \left(b^2 + \frac{2}{3}bf+ \frac{1}{5}f^2 \right)2 \sigma_8 \frac{\partial \sigma_8}{\partial q_i }                 
\end{equation}
with
\begin{equation}
  \frac{\partial f(z)}{q_i} = \frac{\partial \Omega_m(z) }{\partial
    q_i }\bigg( \frac{5}{2g(z)} -\frac{1}{2}\bigg)-  
   \frac{\partial \Omega_m(z ) }{\partial q_i } - 
  \frac{5}{2}\frac{\Omega_m(z)}{g^2(z)}  \frac{\partial g(z) }{\partial q_i },
  \label{fder}
\end{equation}
\begin{equation}
\frac{\partial \sigma_8(z)}{\partial q_i} = \dfrac{\sigma_{8,0} }{1 + z }  \frac{\partial g(z) }{\partial q_i };
\label{sigmader}
\end{equation}
where 
\begin{equation}\begin{split}
 & \frac{\partial g(z) }{\partial q_i } = \frac{5}{2(1+z)^2}\bigg(\frac{\partial \Omega_m(z) }{\partial q_i } 
  H^3(z) +  3H^2(z) \frac{\partial H(z) }{\partial q_i }  \Omega_m(z) \bigg)\cdot
   \\ & \int_z^\infty\mathrm{d}z'\frac{ (1+z')}{H^3(z')} - \frac{15 \Omega_{m}(z)H^3(z)}{2(1+z)^2}\int_z^\infty\mathrm{d}z'\frac{(1+z')}{H^4(z')}\frac{\partial H(z')}{\partial q_i}
\end{split}\end{equation}
\begin{equation}\begin{split}
 \frac{\partial H(z)}{\partial q_i} = H_0 \frac{1}{2} \left( \frac{\Omega_{m,0} (1+z)^3}{\Omega_m(z) }\right)^{-1/2}\frac{(1+z)^3 }{\Omega_m(z) }  \\ \cdot 
 \left( \frac{\partial \Omega_{m,0} }{\partial q_i} - \frac{\Omega_{m,0}}{\Omega_m}\frac{\partial \Omega_{m}( z ) }{\partial q_i} \right),
 \end{split}
\end{equation}
with
\begin{equation}
\begin{split}
\frac{\partial\Omega_m}{\partial q_0}& =  \Omega_{m,fid}(z),\\
\frac{\partial\Omega_m}{\partial q_1}& =  \Omega_{m,\rm fid}(z)y(z),\\
\frac{\partial\Omega_m}{\partial q_2}& =   \Omega_{m,\rm fid}(z)\frac{y^2(z)}{2}.\\
\end{split}
\end{equation}
Note that in the equations above  \rewcom{the $P(k)$ term has been factored out};  all the terms are evaluated at $q_0 = 1$ since we are ignoring the weights dependence on cosmology.

Similarly for the quadrupole and for the hexadecapole
\begin{equation}\begin{split}
w_{2, q_i } &=  \left(\frac{4}{3} b + \frac{8}{7} f \right)\sigma_8^2 \frac{\partial f}{\partial q_i} +  \left(\frac{4}{3} bf + \frac{4}{7} f^2  \right)2\sigma_8 \frac{\partial \sigma_8}{\partial q_i } ,        \\
&\\
w_{4,i} &= \frac{16}{35}  f \sigma_8^2   \frac{\partial f}{\partial q_i}+    \frac{8}{35} f^2  2 \sigma_8\frac{\partial \sigma_8}{\partial q_i }.
\end{split}\end{equation}

\begin{figure}
\centering
\includegraphics[scale=0.75]{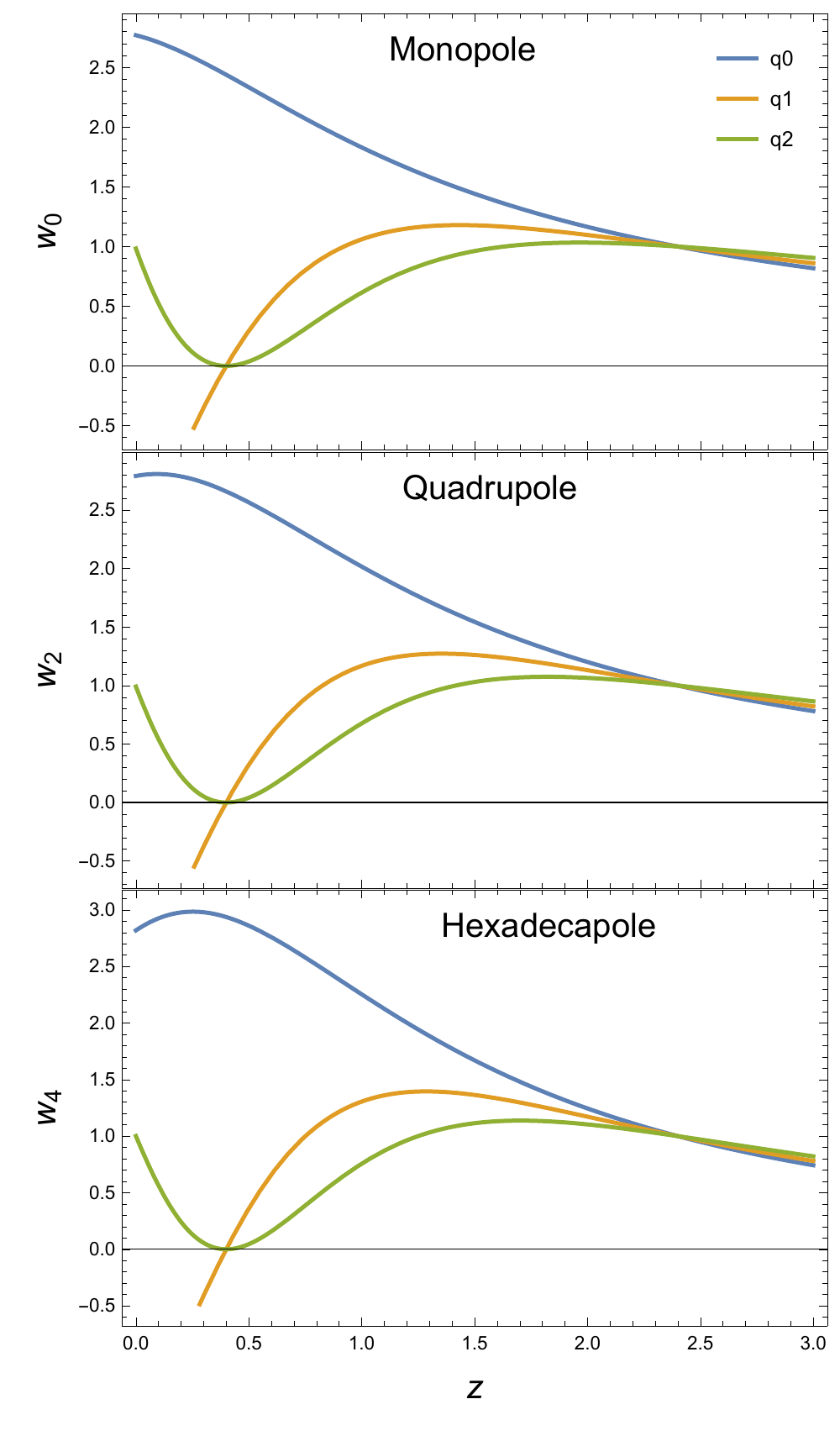}
\caption{The weights for the monopole ($w_0$) , quadrupole ($w_2$) and
  hexadecapole ($w_4$) with respect to the $q_i$ parameters: blue lines
  indicate the weight with respect to $q_0$, orange lines indicate
  the weight with respect to $q_1$ and the green lines the weight with
  respect to $q_2$.  These weights assume a fiducial bias evolving as
  $ b = \sqrt{1+z} $ and were calculated for RSD measurements assuming that the bias is known, as described in Sec. \ref{allinfoweights}.  }
\label{mplot}
\label{hplot}
\label{qplot}
\end{figure}

Figure \ref{mplot} shows the set of weights for the monopole, quadrupole
and hexadecapole, with a convenient normalization. All the plot are generated considering a $\Lambda$CDM  model with $\Omega_{m,0} = 0.31$ as the fiducial cosmology.
We explore a wide redshift range to see  general trends, as if we are analysing data from a range of surveys. We fix a pivot redshift in $z_p = 0.4$.  All
three weights with respect to parameters $q_0$ (blue lines) show a
peak at redshift $z \sim 0.1-0.2$; this is due to
$\partial \Omega_m / \partial q_i$ term which rapidly grows until about
$z \sim 2$ and then tends to a constant. The peak corresponds to the
$\Omega_m \sim \Omega_\Lambda$ epoch: the weights aim to highlight the
deviations from the fiducial cosmology $\Lambda \rm CDM $, and therefore
peak approximately in the range of the equivalence between matter and
$\Lambda$.  At higher redshifts the weights decrease due to the
decreasing dependence of $\Omega_m(z)$ on $f$ and $\sigma_8$.

The weights about the \textit{slope parameters}, $q_1$ (orange line),
 rapidly grow at low redshift driven by $\partial \Omega_m / \partial q_1 $ and
$\partial P_\ell / \partial \sigma_8$, and then start decreasing as
$\partial f / \partial \Omega_m$ and $\partial P_\ell / \Omega_m $
dominate. We see that they pickup small differences about the peak due
to the different dependencies on $P_\ell$.

The green lines displays the weights with respect to the second order
parameter $q_2$: they are similar, with a minimum about $z $
$\sim0.5 $: this difference with respect to $q_0$ and $q_1$ is due to
the $\partial \Omega_m / \partial q_2$ term, which starts decreasing with $z$
until about $z\sim0.4$ and then slowly increases.  Comparing the
monopole, quadrupole and hexadecapole weights we see that the weights
behave in a similar way for all three statistics; the hexadecapole weights
 show a faster decrease for all three parameters,
due to the absence of the bias  dependence. However the
differences with monopole and quadrupole are small, confirming our
assumption that the bias choice does not drastically change the
weights in the region of interest.

  \subsubsection{  Application of the method}
\rewcom{%
We have derived a set of weights that compress  the information available in the power spectrum across  a range of redshifts.
In practice, to apply the method, we weight galaxies, assuming  ${\bf w}_{\rm gal}=\sqrt{{\bf w}_{\delta^2}}$, 
to obtain a set of monopole, quadrupole and hexadecapole for each set of weights. }

\rewcom{If we were only interested in a single parameter, (e.g. $q_0$) and we thought all the information came from the monopole, we would measure  the weighted $P_{w,0}$ by applying the $w_{0,q_0}$ to each galaxy; we would then fit $q_0$ by comparing the data with the theoretical prediction for the monopole, weighted at different redshifts as
 \be
P_{0, w_{0,q_0} \rm model}(k)= \int dz\; P_0(k,z ) \cdot  w_{0,q_0}(z).
\label{thmo}
 \ee
 Where  the $P_0(k,z, q_0) $ corresponds to the monopole prediction, e.g Eq.\ref{m} and we have ignored the window effects. If we further assume the simple linear model for RSD, \citep{kaiser1987},
 \be
   P_{0}(k, z) = \left(b^2 +\frac{2}{3} b f(\Omega_m(q_0, z)) +\frac{1}{5}
    f(\Omega_m(q_0, z))^2\right)P(k);
 \ee
 we can express $f$ in terms of $\Omega_m(q_0, z)$ according to Eq.  \ref{eq:ff}. }

\rewcom{In order to simultaneously measure all three $q_i$ parameters, we measure each multipole weighted to be optimal for each  $q_i $ parameters, i.e. we weight galaxies with the different $w_{i, q_j} $ functions and we build a data vector $\Pi$ as, 
\be
 \Pi^T =( P_{0, w_0,q_0}, P_{0, w_0,q_1}, P_{0, w_0,q_2} \; ...  \; P_{4, w_4,q_2})^{\rm T}.
\ee  
Note that each weighted multipole $P_{i, w_{i, q_j}}$ provides a particular piece of information about  $\Omega_m(z)$ that optimizes the measurement of each $q_i$. 
 We constrain the three $q_i$  by jointly fitting  from the data-vector $\Pi$  compared with a  $\Pi_{\rm model}$.  In practice we assume a Gaussian likelihood and minimize 
\be
\chi^2 \propto ( \Pi -  \Pi_{\rm model} )^{\rm T} C^{-1} (\Pi -  \Pi_{\rm model} );
\ee
where each $P_{i, w_{i, q_j}}$ inside $\Pi_{\rm model}$ is modeled as in Eq. \ref{thmo}. The $C^{-1}$ term corresponds to the joint covariance matrix. }
\subsubsection{The dependence on the fiducial bias model}  \label{biaspart}

We now  test the robustness of the set of weights for $\Omega_m$, presented in \ref{allinfoweights}, with respect to the bias model.
To do this we compute sets of weights from different choices of $b(z)$. 
We first derive the set of weights presented in Sec. \ref{redweigh}, parametrized with respect to $\Omega_m$, fixing a constant bias, $b = 1.024$, which is our fiducial value  at  $z =0.45$, then we repeat for $b = 1/D_{\rm fid}(z)$.

\begin{figure}
\centering
\includegraphics[scale=0.75]{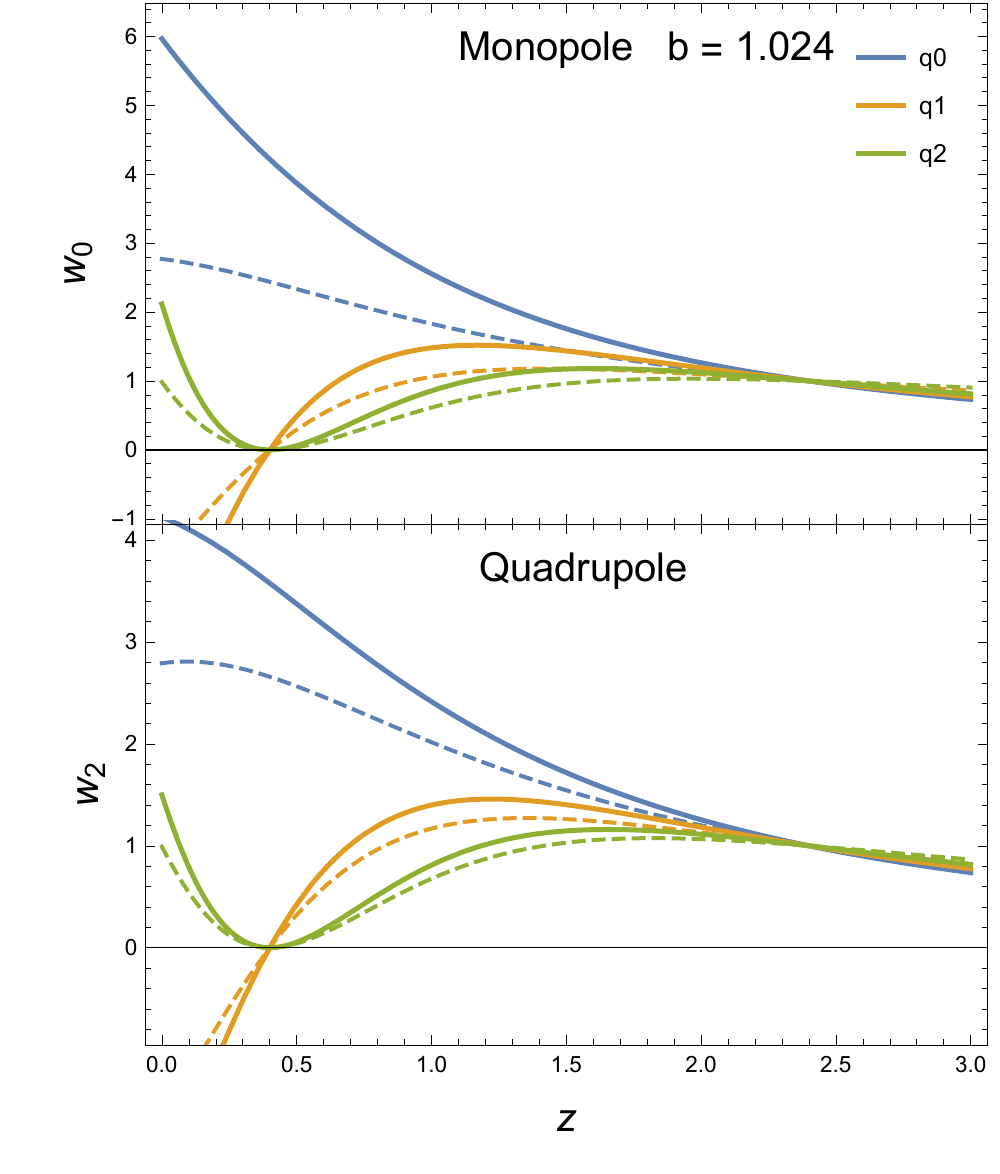}
\caption{Solid lines as Fig. \ref{hplot}, but assuming a constant bias  $b = 1.024$.
We compare this set of weights with the results presented in Fig \ref{hplot}, (dashed lines), obtained assuming a bias evolving as $ b = \sqrt{1+z} $.   }  
\label{cbmplot}
\end{figure}

\begin{figure}
\centering
\includegraphics[scale=0.75]{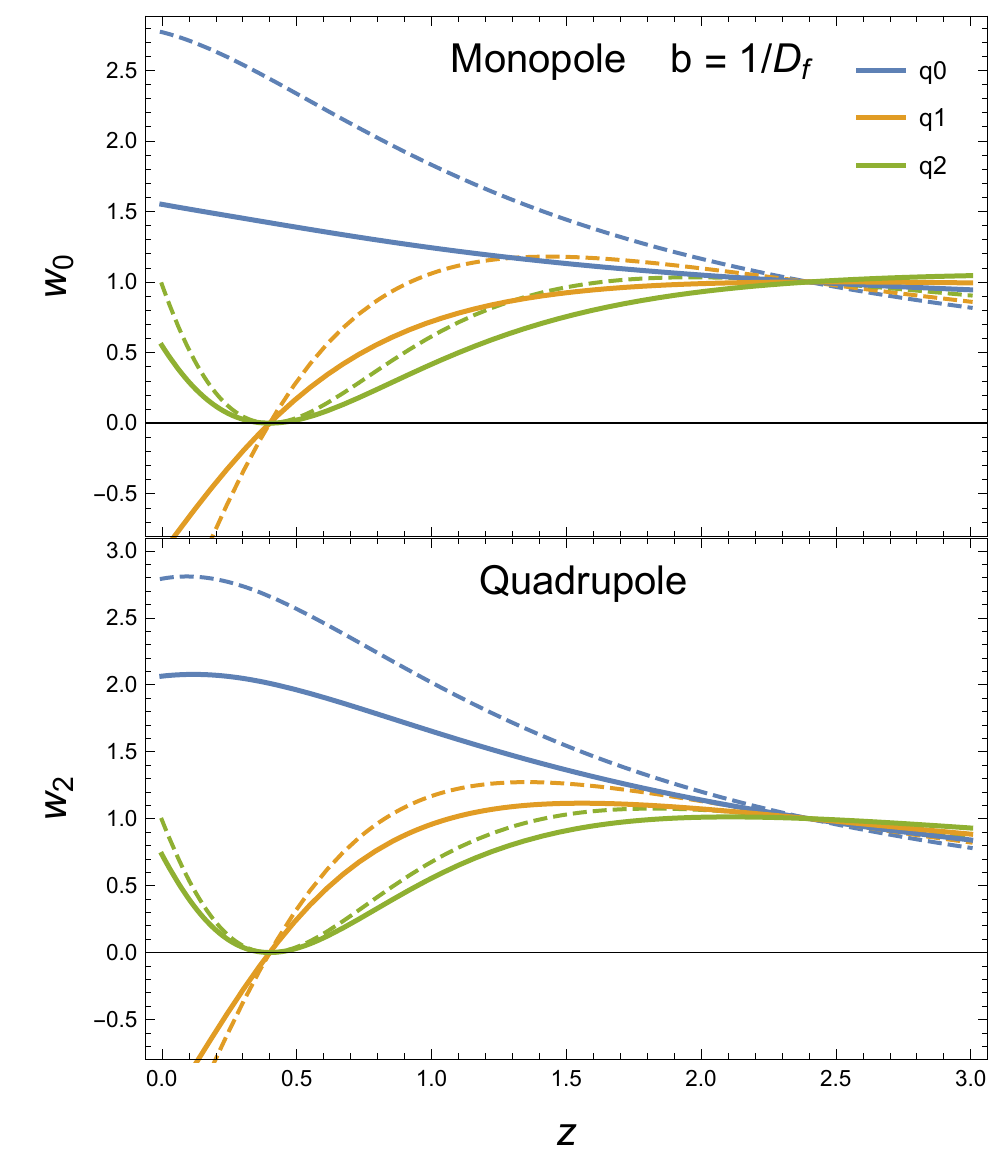}
\caption{Solid lines as Fig. \ref{hplot}, but assuming a fiducial  bias evolving with redshift as $b = 1/D_f(z)$.  We compare this set of weights with the results presented in Fig \ref{hplot}, (dashed lines), obtained assuming a bias evolving as $ b = \sqrt{1+z} $.  }
\label{gbmplot}
\end{figure}

Figures \ref{cbmplot},  \ref{gbmplot} show that the  behaviour  of the weights with redshift is similar to previous results and there are no  significant differences in the shapes. As expected the differences are more visible in the monopole (top panel) than in the quadrupole (bottom panel)  since the former is more sensitive to galaxy bias.  We exclude the weights for the hexadecapole since it does not depend on galaxy bias.

\subsection{Optimal weights to measure $\Omega_m(z)$ with unknown bias}  \label{sec:weights_APfixed_fsig8}


\label{fsig8}
RSD measurements constrain the product of the two key parameters $f$ and $\sigma_8$ and it is common to consider  a single measurement of $[f \sigma_8] $, marginalising over an unknown bias.
Therefore we  present a set of weights that matches the philosophy of current RSD measurements: 
 we consider the term $[b \sigma_8] $ to be \textit{independent} from $[f\sigma_8]$  since we marginalize over the bias.
%
%
Considering e.g. the monopole
\begin{equation}
P_0 = \bigg(  [b \sigma_8]^2 + \frac{2}{3} [b \sigma_8] [f\sigma_8](z) +\frac{1}{5} [f\sigma_8]^2(z) \bigg)P(k)/(\sigma_8^2)
\end{equation}
for unknown bias the dependence on the $q_i$ parameters is only through $[f\sigma_8]$.  
We derive the set of weight  by taking the derivative of $P_0$, $P_2$, $P_4$  with respect to $q_1$, $q_2$, $q_3$, 
 \begin{equation}
w_{0, q_i } \equiv \left( \frac{2}{3} [b \sigma_8 ]  + \frac{2}{5} [f\sigma_8](z) \right)  \frac{\partial [f\sigma_8]}{\partial q_i} (z),
\end{equation}
 \begin{equation}
w_{2, q_i }  \equiv \left( \frac{4}{3}[ b \sigma_8]   + \frac{8}{7} [f\sigma_8](z) \right)\frac{\partial [f\sigma_8]}{\partial q_i} (z),
\end{equation}
 \begin{equation}
w_{4, q_i } \equiv \left( \frac{16}{35} [f\sigma_8](z) \right) \frac{\partial [f\sigma_8]}{\partial q_i} (z),
\end{equation}
where the derivatives   $\partial [f\sigma_8]/\partial q_i (z)$ are obtained  using Eq. \ref{fder} and \ref{sigmader}.
\begin{figure}
\centering
\includegraphics[scale=0.75]{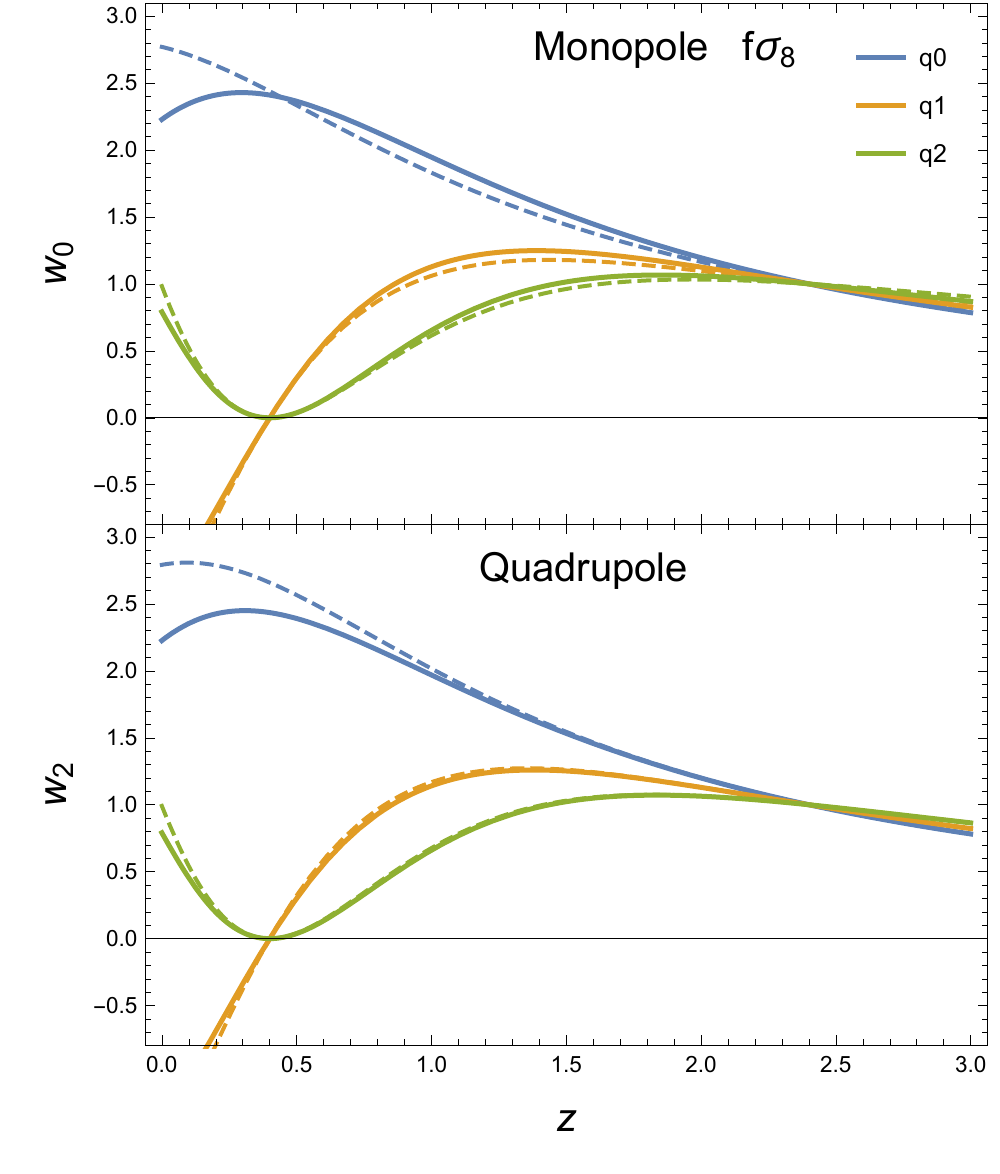}
\caption{
Solid lines show weights for the monopole ($w_0$)  quadrupole ($w_2$) and  hexadecapole ($w_4$) , with respect to the $q_i$ parameters ignoring the information in $b \sigma_8$ term, as described in Sec. \ref{fsig8}. 
These are compared with the weights presented in Fig. \ref{hplot} (dashed lines), which assume the bias is known.
}
\label{fsmplot}
\end{figure}

Figure \ref{fsmplot},  shows the set of weights for the monopole, quadrupole and hexadecapole parametrized with respect to $q_0$, $q_1$, $q_2$, when ignoring the information contained in $[b \sigma_8]$, conveniently normalised. We compare them with the weights derived in \ref{allinfoweights}, presented in Fig \ref{hplot}, (dashed lines). 
The main difference between the two set of weights lies on the assumptions we make for galaxy bias: if we are setting it as completely unknown, considering only the information contained in $[f\sigma_8]$ or if we are including $[b\sigma_8]$ term, constraining $b(z)$ to a fiducial model; however the plots show a very similar behaviour between the two cases, it is clear then  that the tangential modes do not play a large role in determining optimal weights.
\subsection{Optimal weights to measure bias}   \label{sec:weights_APfixed_bias}
For completeness we will show how to derive weights that optimally measure the evolution of
the bias parameter around the fiducial model. 
  
In an analogous manner to Eq. \ref{espomega} we model $[b\sigma_8](z)$ as an expansion about
a fiducial model $[b\sigma_8]_{\rm fid}$:
\begin{equation}
\dfrac{[b\sigma_8](z) }{[b\sigma_8]_{\rm fid} (z) } = \eta_0 \left(1 + \eta_1 x + \dfrac{1}{2}\eta_2 x^2 \right)
\end{equation}
about a pivot redshift $z_p$, where $1 + x \equiv \frac{[b\sigma_8]_{\rm fid}(z)}{[b\sigma_8]_{\rm fid}(z_p) }$.
 \\\\
The $\eta $ parameters correspond to the derivative 0,1,2 of $[b\sigma_8](z)$ relation evaluated at $z_p$.

%
%
%

In analogy with measuring $q_i$, we can derive a set of weights that
optimally estimate the bias-redshift through the $\eta_j$
parameters,
\begin{equation}
w_{\ell, \eta_j} = \frac{\partial P_\ell }{ \partial [b\sigma_8]} \frac{\partial [b\sigma_8]}{\eta_j} 
\end{equation}
This set of weights can be applied instead of the set of weights
with respect to $\Omega_m$ in case we want to measure deviations from
the fiducial model chosen for the bias.
We do not plot these weights for simplicity but include the derivation to show how they could be calculated.

\section{Redshift weighting assuming unknown distance-redshift relation}
\label{biascompar}
Our previous results provide an optimal scheme specific for RSD measurements; as pointed out in the introduction it would be very useful to optimize at the same time geometric measurements and thus enable measurements that include all the parameters both for computational costs either for accuracy of the results. 
%
%
%
%
%
%
An optimal weighted scheme for BAO measurements has been recently presented in \cite{zhu2014} where the authors describe a weighting scheme parametrized with respect to the distance-redshift relation, 
including the AP effect modelled as
\begin{equation}\begin{split}
k_\perp &\rightarrow \alpha^{-1} (1+ \epsilon) k_\perp,\\
k_\parallel &\rightarrow \alpha^{-1} (1+ \epsilon)^{-2} k_\parallel,
\end{split}
\end{equation}
with parameters $\alpha$,  for isotropic deformation and  $\epsilon$, for anisotropic.
The method optimises only BAO measurements, constraining the covariance matrix at BAO scales and ignoring the growth parameters. 


In this section we will account for both distortions due to peculiar velocities  and distortions due to incorrect choice of geometry described by Alcock-Paczy\'nski effect.
We still use a parametrization of $\Omega_m(z)$ to define deviations from our fiducial model as described in Eq. \ref{espomega}.
\subsection{Modelling AP and RSD in the observed $P(k)$ }\label{genpol}
We denote $k^{t}$  and $k$ the true and observed coordinates respectively, then assuming an incorrect geometry
transforms the coordinates 
\begin{equation}\begin{split}
k^{t} &= \frac{k}{\alpha_\perp } \left[ 1 + \mu^2 \left( \frac{\alpha_\perp^2}{\alpha_\parallel^2} - 1 \right) \right]^{1/2},\\
\mu^{t} &=  \mu \frac{\alpha_\perp}{\alpha_\parallel}\left[ 1 + \mu^2 \left( \frac{\alpha_\perp^2}{\alpha_\parallel^2} - 1 \right) \right]^{-1/2},
\label{trans}
\end{split}
\end{equation}
with
 $\alpha_\parallel $ defined as the ratio between the observed and the true Hubble parameter, $H(z)/ H_t(z)$ and $\alpha_\perp$ defined  as ratio between the true and the observed angular diameter distance $D_{A,t}(z)$ / $D_A(z)$, e.g. \citep{ballinger1996};
the multipoles at the observed $k$ are related to the Power Spectrum at $k^t$, through
\begin{equation}
P_\ell (k)= \frac{2 \ell +1 }{ 2 } \int_{-1}^{1} d \mu P(k^t, \mu^t) \mathcal{L}_\ell (\mu)
\label{uml}
\end{equation}
For linear redshift space distortion, \cite{kaiser1987}, inserting the transformation of the coordinates  given by Eq. \ref{trans}, the galaxy power spectrum  at the true wavenumber  is
\begin{equation}\begin{split}
P^s(k^t, \mu^t) = \dfrac{1}{\alpha_\perp^2 \alpha_\parallel} P\left[  \frac{k}{\alpha_\perp } \left[ 1 + \mu^2 \left( \frac{\alpha_\perp^2}{\alpha_\parallel^2} - 1 \right) \right]^{1/2}  \right]&\\
\left[ 1 + \mu^2 \left( \frac{\alpha_\perp^2}{\alpha_\parallel^2} - 1 \right) \right]^{-2}  \left\{ 1 + \mu^2 \left[ (\beta +1)\frac{\alpha_\perp^2}{\alpha_\parallel^2} - 1 \right] \right\}^{2}&.
\label{pexp}
\end{split}
\end{equation}
We use the notation $\beta \equiv f/b$ for simplicity with equations.
%
 We expand at first order $P$ in the right side of Eq. \ref{pexp}, in order to get analytical derivatives with respect to the expansion parameters.  We have tested numerically that this approximation does not influence our conclusions. 
Introducing  $\varphi \equiv \frac{\alpha_\perp^2}{\alpha_\parallel^2} -1 $, we can expand  the right side of Eq. \ref{pexp} to first order about  $(\alpha_\perp , \varphi )= (1, 0) $,
using
 \begin{equation}\begin{split}
& P\left[  \frac{k}{\alpha_\perp } \left( 1 + \mu^2 \varphi \right)^{1/2}  \right] \approx 
P(k) + 
\\&(\alpha_\perp -1 ) \left. \dfrac{\partial P}{\partial k  } \dfrac{\partial k }{\partial k^t}  \dfrac{\partial k^t }{\partial \alpha_\perp} \right|_{\begin{subarray}{c}{\alpha_\perp = 1}\\
    {\;\;\;\varphi = 0}\end{subarray}} 
+\varphi \left. \dfrac{\partial P}{\partial k  } \dfrac{\partial k }{\partial k^t}  \dfrac{\partial k^t }{\partial \varphi}\right|_{\begin{subarray}{c}{\alpha_\perp = 1}\\
    {\;\;\;\varphi = 0}\end{subarray} } 
\end{split}
\end{equation}
Substituting in Eq. \ref{pexp} and then in Eq. \ref{uml}, we obtain models of the multipoles accounting for both RSD and AP effects to be
\begin{equation}\begin{split}
&P_{\ell} (k)= \dfrac{2 \ell +1 }{ 2 } \int_{-1}^{1} d \mu  \mathcal{L}_\ell (\mu)\left\lbrace  P(k)\left(1+ \mu^2 \beta \right)^2 + \right.\\
& \varphi\left[\dfrac{1}{2}\dfrac{\partial P}{ \partial \ln k}\mu^2 \left(  1+ \mu^2 \beta \right)^2 + P(k)(-2 \mu^2 )(1+ \beta \mu^2 )^2 +\right.\\
&\left. 2\mu^2(1+ \beta \mu^2)(1+ \beta)P(k) -\dfrac{1}{2}P(k)(1+ \mu^2 \beta)^2  \right] +\\
& \left. (1 - \alpha_\perp) (1 + \mu^2 \beta)^2 \left(\frac{\partial P}{\partial \ln k } + 3 P(k) \right) \right\rbrace 
\end{split}
\end{equation}
In particular it holds  that $P^s(k)$ at linear order is described by the first three multipoles, \cite{ross2015}.
The  monopole, quadrupole and hexadecapole are
\begin{equation}\begin{split}
P_0(k) &=\dfrac{1}{2} \sigma_8^2\bigg\lbrace \left(2 + \dfrac{4}{3} \beta + \dfrac{2}{5} \beta^2 \right)  P(k) +  
\varphi \bigg[ \bigg(\dfrac{1}{3} +  \dfrac{2}{5} \beta+  \dfrac{1}{7}\beta^2  \bigg) \\
& \dfrac{\partial P }{\partial \ln k } -  \left(1 + \dfrac{2}{15} \beta - \dfrac{1}{35}\beta^2 \right)  P(k) \bigg]
+ (1 - \alpha_\perp ) \bigg(2 \\ &+\dfrac{4}{3} \beta + 
\dfrac{2}{5}  \beta^2\bigg)\left(\frac{\partial P}{\partial \ln k } + 3 P(k) \right)   \bigg\rbrace,\\
P_2(k) &=  \dfrac{5}{2} \sigma_8^2\bigg\lbrace \bigg( \dfrac{8}{15} \beta + \dfrac{8}{35} \beta^2 \bigg) P(k)   
+\varphi\bigg[\bigg( \dfrac{2}{15} + \dfrac{8}{35}\beta  \\ & + \dfrac{2}{21}\beta^2 \bigg)\dfrac{\partial P}{\partial \ln k}-\left.\left( \dfrac{4}{21}\beta +\dfrac{4}{105}\beta^2 \right)P(k) 
 \right ] + (1-   \\&  \alpha_\perp ) \left( \dfrac{8}{15}\beta  + \dfrac{8}{35}\beta^2    \right)
 \left. \left(\frac{\partial P}{\partial \ln k } + 3 P(k) \right)\right\rbrace,
\\
P_4(k) &=\dfrac{9}{2} \sigma_8^2 \bigg\lbrace   \dfrac{16}{315} \beta^2 P(k)  
+\varphi\left[\left( \dfrac{16}{315}\beta + \dfrac{8}{231}\beta^2 \right)\dfrac{\partial P}{\partial \ln k} \right.\\
&-\left.\left( \dfrac{32}{315}\beta +\dfrac{24}{385}\beta^2 \right)P(k) 
 \right] + (1- \alpha_\perp ) \dfrac{16}{315}\beta^2      \bigg(\frac{\partial P}{\partial \ln k } + \\
& 3 P(k) \bigg)\bigg\rbrace.
\end{split}
\label{newmonop}
\end{equation}

\subsection{AP and RSD weights derivation, assuming known bias  }
As before, the weights for the power spectrum multipoles, assuming information from both RSD and AP effects, are obtained by taking the derivative of the $P_i$  with respect to the $q_i$ parameters  defined in \ref{sec:dz},
\begin{equation}\begin{split}
w_{\ell, q_i} =  \dfrac{\partial P_\ell }{ \partial q_i } &= \frac{\partial P_\ell }{\partial \varphi} \frac{\partial \varphi}{\partial q_i}  +
\frac{\partial P_\ell }{\partial \alpha_\perp} \frac{\partial \alpha_\perp }{\partial q_i}\\
 &+ 
\frac{\partial P_\ell}{\partial \beta} \frac{\partial \beta}{\partial f}\frac{\partial f}{\partial q_i} +
\frac{\partial P_\ell }{\partial \sigma_8} \frac{\partial \sigma_8 }{\partial q_i}.
\label{sprimweigh}
\end{split}
\end{equation}
 all  the derivatives  are evaluated at the fiducial model. 
 In case of flat universe  we have
%
%
%
\begin{equation} \alpha_\perp(z)
= \frac{\int^z_0 \mathrm{d}z'  1/H(q_i , z')  }{\int^z_0 \mathrm{d}z'' 1/H_{\rm fid}(z'')}
  \end{equation}
  Inserting the definition of $\beta$ and $\varphi$,
\begin{equation}\begin{split}
& \frac{\partial \beta}{\partial f} = \frac{1}{b},\\
&\dfrac{\partial \varphi }{\partial q_i} = - \dfrac{2\alpha_\perp^3}{\alpha_\parallel^3} \bigg( \frac{1}{\alpha_\perp} \frac{\partial \alpha_\parallel}{ \partial q_i} + 
\frac{-\alpha_\parallel}{\alpha^2_\perp} \frac{\partial \alpha_\perp}{ \partial q_i}  \bigg),\\
\end{split} \end{equation}
where
\begin{equation}
\begin{split}
\frac{\partial \alpha_\parallel}{ \partial q_i} &= -\frac{1}{H_{\rm fid}(z)}  \frac{\partial H}{\partial q_i},\\
\frac{\partial \alpha_\perp}{ \partial q_i} &=   = \frac{\int^z_0 \mathrm{d}z'  -\frac{1}{H_{\rm fid}^2( z')}  \frac{\partial H}{\partial q_i}  }{\int^z_0 \mathrm{d}z'' 1/H_{\rm fid}(z'')}.\\
\label{cool}
\end{split}
\end{equation}
In Appendix \ref{append} we present the derivatives of the multipoles $P_i$ with respect to $\varphi$,  $\alpha_\perp$, $f$ and $\sigma_8$. 

The weights have arbitrary normalization but we cannot factor out the scale dependence as we did for the RSD weights since we now have two different $k$-dependent terms $P$ and $dP/d\ln k$. 
However \cite{gongbocit} 
 show that this dependence is very weak. 

\begin{figure}
\centering
\includegraphics[scale=0.76]{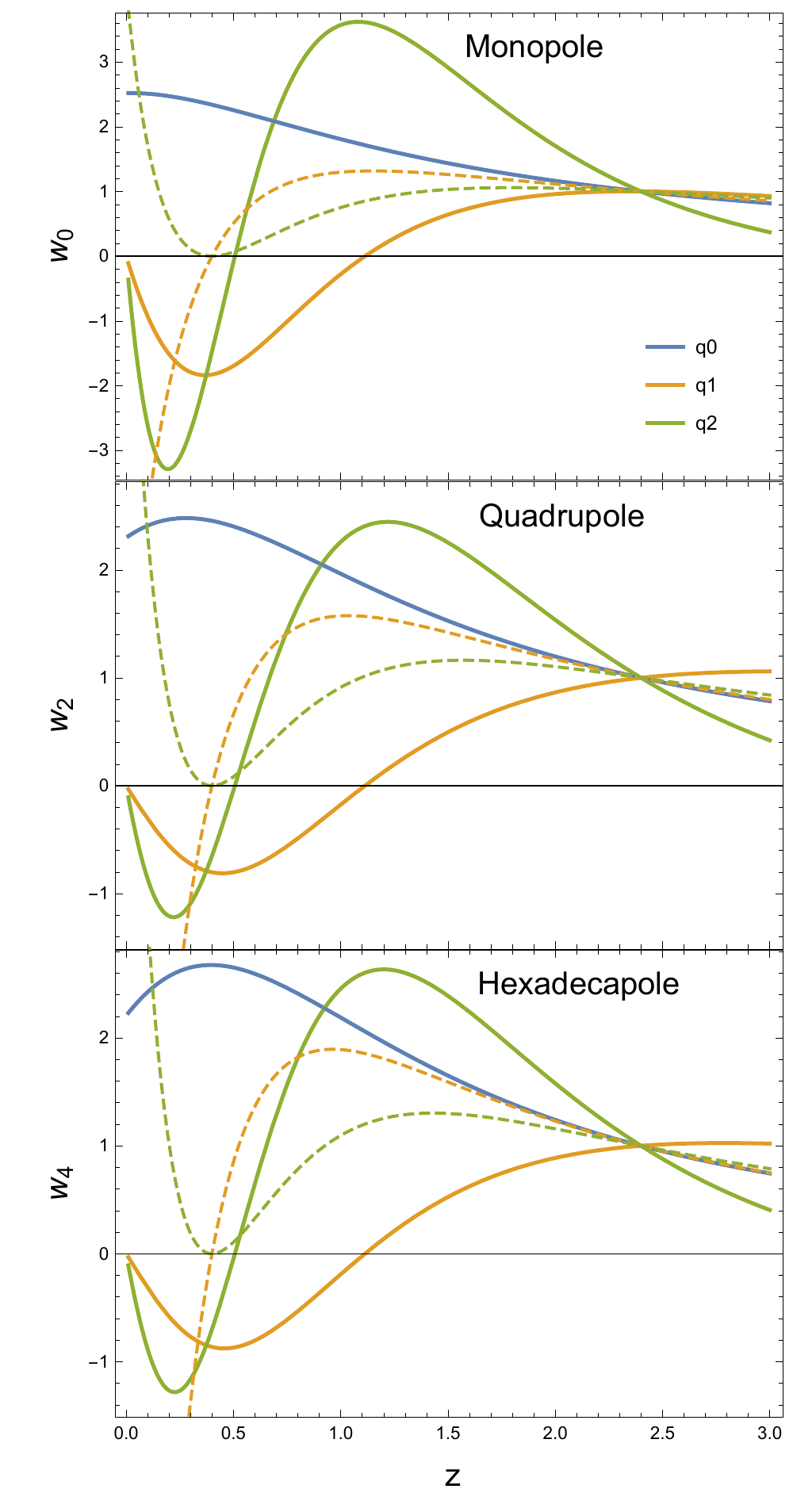}
\caption{The RSD + AP optimal weights for the monopole $w_0$, quadrupole $w_2$, hexadecapole $w_4$. Blue lines indicate the weights with respect to the $q_0$ parameter. Orange lines indicate the weights with respect to the $q_1$ parameter and green lines indicates the weights with respect to the $q_2$ parameter. We compare them with previous results, where the distance-redshift relation was fixed (dashed lines);  the weights $w_{\ell,q_0}$ (blue lines) are equivalent to the previous weights without AP effect since the contribute from $\varphi$ and $\alpha_\perp$ vanish because of $\partial H(q_i)/\partial q_0 =0$. The RSD+AP weights show  enhanced features  due to AP parameters, whose importance increases with redshift.}  
\label{mrsdbaoplot}
\end{figure}

Figure \ref{mrsdbaoplot} shows the weights optimal for RSD and AP measurements evaluated at $k =0.1$ $h\; \rm MPc^{-1}$, for the  monopole, quadrupole and hexadecapole respectively. 
Blue lines indicate the weights with respect to  $q_0$, orange lines with respect to  $q_1$  and green lines with respect to  $q_2$.  
Comparing with the previous result that assumed a known distance-redshift relation (dashed lines), it is possible to see that the behaviour of the three weights  does not change drastically. In general the redshift dependence is stronger including also the AP effect and the new weights show a more enhanced maximum.
Since the contribution from $\varphi$ and $\alpha_\perp$ vanish for $q_0$,
 the weights $w_{i,q_0}$ are equivalent to the previous weights without AP effect. (Fig \ref{mplot}).\\ 

\subsection{AP-RSD weights assuming unknown bias}  
If we now  neglect the information given by $[b\sigma_8]$, as we did for one set of  RSD weights, we substitute $\beta = f/b $, then we change the Eq. \ref{cool} to 
\be
\begin{split}
w_{\ell, q_i} =  \dfrac{\partial P_\ell }{ \partial q_i } &= \frac{\partial P_\ell }{\partial \varphi} \frac{\partial \varphi}{\partial _i}  +
\frac{\partial P_\ell }{\partial \alpha_\perp} \frac{\partial \alpha_\perp }{\partial q_i}\\
 &+ 
\frac{\partial P_\ell}{\partial f \sigma_8} \frac{\partial f \sigma_8 }{\partial q_i},
\label{sprimweigh}
\end{split}
\ee
where we have assumed  that $\partial [b\sigma_8]/\partial q_i = 0$.

 In Sec \ref{redweigh} we showed that there are no  significant differences between  the cases in which $b\sigma_8$ is known and unknown, however, for the reasons  discussed in Sec. 4, they are more consistent with the RSD measurements. We do not plot any new results since the differences are very small. 

\section{Discussion}
\label{discussion}
In the first part of the paper we presented a set of optimal redshift dependent weights for  RSD measurements. These functions allow us to optimally compress the original dataset  while minimising  the error \textit{a priori} provided by the Fisher matrix, on the parameters we want to measure. 
 \rewcom{In contrast to current RSD measurements, which compare the data with the model at a single effective redshift, the weights presented account for the redshift evolution of the cosmological effects,
 since the measurements are now compared with a weighted model covering a range of redshifts. } 
 The method  is based on a particular choice of a relation in redshift we want to measure/investigate, e.g 
modelled as an expansion  about a fiducial model. 
We derived the set of weights that optimally estimates deviations from a $\Lambda C D M $ background, modelled using variations of $\Omega_m(z)$;
we modelled $\Omega_m(z)$ as a polynomial expansion in terms of  parameters $q_i$   about a fixed fiducial models $\Omega_{m,\rm fid}$, then we applied the  compression method as described in \cite{Tegmark97}  and derived the set of weights that optimally estimates the expansion parameters.
 As explained in section \ref{sec:optweights}, our weights generalise the FKP weights: they take in account of the galaxy density distribution through the Covariance Matrix and allow sensitivity to the redshift dependence of the statistics we are measuring. 
 
We started assuming a known distance-redshift relation and considering two different cases: one in which the bias is known and constrained to a fiducial model and one in which the bias is unknown and we marginalize over it. 
We compared them and we showed the consistency between the two scheme: this means that the bias does not play a large role in determining the weights. 
We also  tested our assumptions on the fiducial bias model  by deriving the previous results for two different bias  models. We confirmed also in this case that the weights are not significantly sensitive to $b(z)$.  For completeness we presented a  further set of weights that optimize the measurement of  bias.
%
 %
 
In order to improve future measurements we extended the previous weights to a more general case, when the distance-redshift relation is unknown: we modelled  the observed power spectrum including  the  AP effect;  we introduced two new  distortion parameters $\varphi$, $\alpha_\perp$  in $P(k)$  describing  the AP effect.
%
The set of weights obtained for RSD and  AP measurements is scale dependent through the ratio of $P(k)$ and its logarithmic derivative. This is not a particular issue for future applications since this dependence is predicted to be very weak, moreover 
considering e.g. Camb, \citep{Lewis:2002ah},  computing $P(k)$ model in real space or its derivative will not require high computational time, provided we  apply these weights after calculating the power.  Zhu et al. did not face with this problem since they constrained every quantity on the BAO scale (k $\sim$ $0.1 h Mpc^{-1}$ ). 
We compared the new results with the weights accounting for RSD only at BAO scale showing that the redshift dependence is now increased by the inclusion of the AP parameters. 

In order to  correctly apply the weighting scheme we will need to understand how to combine  with  weights designed to correct for systematic density field distortions.
The derivations presented made a number of assumptions (e.g the fact that the Power Spectrum shape is equal for each redshift slice and that $P$ is Gaussian distributed). It would be interesting to see if the weights change when we relax these assumptions -we leave this for future work.
%
%
%
%
\section*{Acknowledgement}
RR and WJP acknowledge support from the European Research Council through grant {\it Darksurvey}.  WJP is also grateful for support from the UK Science \& Technology Facilities Council through the consolidated grant ST/K0090X/1 and support from the UK Space Agency through grant ST/K00283X/1.
HGM acknowledges the Agence Nationale de la Recherche, as part of the programme Investissements d'avenir under the reference ANR-11-IDEX-0004-02.
GBZ and YW are supported by the Strategic Priority Research Program "The Emergence of Cosmological Structures" of the Chinese Academy of Sciences Grant No. XDB09000000

\def\jnl@style{\it}
\def\aaref@jnl#1{{\jnl@style#1}}

\def\aaref@jnl#1{{\jnl@style#1}}

\def\aj{\aaref@jnl{AJ}}                   
\def\araa{\aaref@jnl{ARA\&A}}             
\def\apj{\aaref@jnl{ApJ}}                 
\def\apjl{\aaref@jnl{ApJ}}                
\def\apjs{\aaref@jnl{ApJS}}               
\def\ao{\aaref@jnl{Appl.~Opt.}}           
\def\apss{\aaref@jnl{Ap\&SS}}             
\def\aap{\aaref@jnl{A\&A}}                
\def\aapr{\aaref@jnl{A\&A~Rev.}}          
\def\aaps{\aaref@jnl{A\&AS}}              
\def\azh{\aaref@jnl{AZh}}                 
\def\baas{\aaref@jnl{BAAS}}               
\def\jrasc{\aaref@jnl{JRASC}}             
\def\memras{\aaref@jnl{MmRAS}}            
\def\mnras{\aaref@jnl{MNRAS}}             
\def\pra{\aaref@jnl{Phys.~Rev.~A}}        
\def\prb{\aaref@jnl{Phys.~Rev.~B}}        
\def\prc{\aaref@jnl{Phys.~Rev.~C}}        
\def\prd{\aaref@jnl{Phys.~Rev.~D}}        
\def\pre{\aaref@jnl{Phys.~Rev.~E}}        
\def\prl{\aaref@jnl{Phys.~Rev.~Lett.}}    
\def\pasp{\aaref@jnl{PASP}}               
\def\pasj{\aaref@jnl{PASJ}}               
\def\qjras{\aaref@jnl{QJRAS}}             
\def\skytel{\aaref@jnl{S\&T}}             
\def\solphys{\aaref@jnl{Sol.~Phys.}}      
\def\sovast{\aaref@jnl{Soviet~Ast.}}      
\def\ssr{\aaref@jnl{Space~Sci.~Rev.}}     
\def\zap{\aaref@jnl{ZAp}}                 
\def\nat{\aaref@jnl{Nature}}              
\def\iaucirc{\aaref@jnl{IAU~Circ.}}       
\def\aplett{\aaref@jnl{Astrophys.~Lett.}} 
\def\apspr{\aaref@jnl{Astrophys.~Space~Phys.~Res.}}
\def\bain{\aaref@jnl{Bull.~Astron.~Inst.~Netherlands}} 
\def\fcp{\aaref@jnl{Fund.~Cosmic~Phys.}}  
\def\gca{\aaref@jnl{Geochim.~Cosmochim.~Acta}}   
\def\grl{\aaref@jnl{Geophys.~Res.~Lett.}} 
\def\jcp{\aaref@jnl{J.~Chem.~Phys.}}      
\def\jgr{\aaref@jnl{J.~Geophys.~Res.}}    
\def\jqsrt{\aaref@jnl{J.~Quant.~Spec.~Radiat.~Transf.}}
\def\memsai{\aaref@jnl{Mem.~Soc.~Astron.~Italiana}}
\def\nphysa{\aaref@jnl{Nucl.~Phys.~A}}   
\def\physrep{\aaref@jnl{Phys.~Rep.}}   
\def\physscr{\aaref@jnl{Phys.~Scr}}   
\def\planss{\aaref@jnl{Planet.~Space~Sci.}}   
\def\procspie{\aaref@jnl{Proc.~SPIE}}   
\def\jcap{\aaref@jnl{J. Cosmology Astropart. Phys.}}

\let\astap=\aap
\let\apjlett=\apjl
\let\apjsupp=\apjs
\let\applopt=\ao

\newcommand{\etal}{et al.\ }

\newcommand{\mpc}{\, {\rm Mpc}}
\newcommand{\kpc}{\, {\rm kpc}}
\newcommand{\hmpc}{\, h^{-1} \mpc}
\newcommand{\ihmpc}{\, h\, {\rm Mpc}^{-1}}
\newcommand{\ikms}{\, {\rm s\, km}^{-1}}
\newcommand{\kms}{\, {\rm km\, s}^{-1}}
\newcommand{\hkpc}{\, h^{-1} \kpc}
\newcommand{\lya}{Ly$\alpha$\ }
\newcommand{\lyb}{Lyman-$\beta$\ }
\newcommand{\lyaf}{Ly$\alpha$ forest}
\newcommand{\lr}{\lambda_{{\rm rest}}}
\newcommand{\bF}{\bar{F}}
\newcommand{\bS}{\bar{S}}
\newcommand{\bC}{\bar{C}}
\newcommand{\bB}{\bar{B}}
\newcommand{\vdF}{{\mathbf \delta_F}}
\newcommand{\vdS}{{\mathbf \delta_S}}
\newcommand{\vdf}{{\mathbf \delta_f}}
\newcommand{\vdn}{{\mathbf \delta_n}}
\newcommand{\vdC}{{\mathbf \delta_C}}
\newcommand{\vdX}{{\mathbf \delta_X}}
\newcommand{\xrei}{x_{rei}}
\newcommand{\lrmin}{\lambda_{{\rm rest, min}}}
\newcommand{\lrmax}{\lambda_{{\rm rest, max}}}
\newcommand{\lmin}{\lambda_{{\rm min}}}
\newcommand{\lmax}{\lambda_{{\rm max}}}
\newcommand{\hi}{\mbox{H\,{\scriptsize I}\ }}
\newcommand{\heii}{\mbox{He\,{\scriptsize II}\ }}
\newcommand{\vp}{\mathbf{p}}
\newcommand{\vq}{\mathbf{q}}
\newcommand{\vxperp}{\mathbf{x_\perp}}
\newcommand{\vkperp}{\mathbf{k_\perp}}
\newcommand{\vrperp}{\mathbf{r_\perp}}
\newcommand{\vx}{\mathbf{x}}
\newcommand{\vy}{\mathbf{y}}
\newcommand{\vk}{\mathbf{k}}
\newcommand{\vR}{\mathbf{r}}
\newcommand{\tdtwo}{\tilde{b}_{\delta^2}}
\newcommand{\tstwo}{\tilde{b}_{s^2}}
\newcommand{\tbthree}{\tilde{b}_3}
\newcommand{\tadtwo}{\tilde{a}_{\delta^2}}
\newcommand{\tastwo}{\tilde{a}_{s^2}}
\newcommand{\tabthree}{\tilde{a}_3}
\newcommand{\vnabla}{\mathbf{\nabla}}
\newcommand{\tpsi}{\tilde{\psi}}
\newcommand{\vv}{\mathbf{v}}
\newcommand{\fnl}{{f_{\rm NL}}}
\newcommand{\tfnl}{{\tilde{f}_{\rm NL}}}
\newcommand{\gnl}{g_{\rm NL}}
\newcommand{\orderfour}{\mathcal{O}\left(\delta_1^4\right)}
\newcommand{\SDSSPF}{\cite{2006ApJS..163...80M}}
\newcommand{\PF}{$P_F^{\rm 1D}(k_\parallel,z)$}
\newcommand\ionalt[2]{#1$\;${\scriptsize \uppercase\expandafter{\romannumeral #2}}}%
\newcommand{\vxone}{\mathbf{x_1}}
\newcommand{\vxtwo}{\mathbf{x_2}}
\newcommand{\vRot}{\mathbf{r_{12}}}
\newcommand{\cm}{\, {\rm cm}}

\bibliographystyle{mn2e}
\bibliography{draft.bib}

\appendix
\label{append}
\section{Derivative of the multipoles with respect to $\varphi$, $\alpha_\perp$,  $\beta$, $\sigma_8$. }
We here present the derivatives of the generalised multipoles computed in Sec. \ref{genpol}, with respect to the redshift dependent parameters. 
\begin{equation}\begin{split}
\frac{\partial P_0}{ \partial \varphi} =&  \frac{1}{2}\sigma_8^2\bigg[\bigg(\frac{1}{3 } + \frac{2}{5} \beta  + \frac{1}{7} \beta^2 \bigg) \frac{\partial P}{\partial \ln k} - \bigg(1 + \dfrac{2}{15} \beta -\\
& \dfrac{1}{35}\beta^2 \bigg)  P(k) \bigg] \\
\frac{\partial P_0}{ \partial \alpha_\perp} =& - \frac{1}{2} \sigma_8^2 \bigg(2 +\dfrac{4}{3} \beta + 
\dfrac{2}{5}  \beta^2\bigg)\bigg(\frac{\partial P}{\partial \ln k } + 3 P(k) \bigg)  \\
\frac{\partial P_0}{ \partial \beta} =& \frac{1}{2}\sigma_8^2 \bigg\{  \bigg(  \frac{4}{3} + \frac{4}{5}\beta   \bigg)P(k)  \\
\frac{\partial P_0}{\partial \sigma_8} =&\sigma_8  \bigg(2 + \dfrac{4}{3} \beta + \dfrac{2}{5} \beta^2 \bigg)  P(k)   
\label{derivgen}
\end{split}
\end{equation}

\begin{equation}
\begin{split}
\frac{\partial P_2}{\partial \varphi} =& \frac{5}{2}\sigma_8^2 \bigg[ \bigg( \dfrac{2}{15} + \dfrac{8}{35}\beta   + \dfrac{2}{21}\beta^2 \bigg)\dfrac{\partial P}{\partial \ln k}-\bigg( \dfrac{4}{21}\beta +\dfrac{4}{105}\beta^2 \\ &\bigg)P(k) 
  \bigg]\\
\frac{\partial P_2}{\partial \alpha_\perp } =& -\frac{5}{2}\sigma_8^2 \bigg( \dfrac{8}{15}\beta  + \dfrac{8}{35}\beta^2    \bigg)
  \bigg(\frac{\partial P}{\partial \ln k } + 3 P(k)  \bigg)\\
\frac{\partial P_2}{\partial \beta} =& \frac{5}{2}\sigma_8^2  \bigg( \frac{8}{15} + \frac{16}{35}  \beta  \bigg)P(k)   \\
\frac{\partial P_2}{\partial \sigma_8} =& 5 \sigma_8  \bigg( \dfrac{8}{15} \beta + \dfrac{8}{35} \beta^2 \bigg) P(k)  
\end{split}
\end{equation}

\begin{equation}
\begin{split}
\frac{\partial P_4 }{\partial \varphi } =& \dfrac{9}{2} \sigma_8^2  \bigg[\left( \dfrac{16}{315}\beta + \dfrac{8}{231}\beta^2 \right)\dfrac{\partial P}{\partial \ln k} 
-\left( \dfrac{32}{315}\beta +\dfrac{24}{385}\beta^2 \right)\\P(k) \bigg]\\
\frac{\partial P_4 }{\partial \alpha_\perp }  =&-\dfrac{9}{2} \sigma_8^2 \bigg\lbrace  \dfrac{16}{315}\beta^2      \bigg(\frac{\partial P}{\partial \ln k } +  3 P(k) \bigg) \bigg\rbrace\\
\frac{\partial P_4 }{\partial \beta }=&\dfrac{9}{2} \sigma_8^2  \dfrac{32}{315} \beta P(k)  \\
\frac{\partial P_4 }{\partial \sigma_8 } =&9\sigma_8     \dfrac{16}{315} \beta^2 P(k)  
\end{split}
\end{equation}
In case we wish to neglect the $[b \sigma_8]$ term, we need to substitute the derivatives w.r.t. $\beta$ and $\sigma_8$, with the derivatives  $\partial P_i /\partial [f \sigma_8 ]$.
\end{document}